\documentclass{article}
\usepackage{amsmath}
\usepackage{color}
\usepackage{amsmath, amssymb}

\DeclareMathOperator*{\regprod}{\mathchoice%
{\ooalign{\hbox{$\displaystyle\prod$}\crcr\hbox{$\displaystyle\coprod$}}}
{\ooalign{\hbox{$\textstyle\prod$}\crcr\hbox{$\textstyle\coprod$}}}
{\ooalign{\hbox{$\scriptstyle\prod$}\crcr\hbox{$\scriptstyle\coprod$}}}
{\ooalign{\hbox{$\scriptscriptstyle\prod$}\crcr\hbox{$\scriptscriptstyle\coprod$}}}
}
\newtheorem{thm}{Theorem}[section]

\newtheorem{cor}[thm]{Corollary}
\newtheorem{pro}[thm]{Proposition}

\newtheorem{deff}{Definition}[section]

\newcommand{\ket}[1]{|#1 \rangle}
\newcommand{\bra}[1]{\langle #1|}

\newcommand{\bec}[1]{\mbox{\boldmath $#1$}}

\begin{document}

\title{{\bf Grover algorithm and absolute zeta functions}
\vspace{15mm}}

\author{Jir\^o AKAHORI$^{1}$, $\quad$ Kazuki HORITA$^{2,\ast}$, $\quad$ Norio KONNO$^{3}$, 
\\
\\
Rikuki OKAMOTO$^{4}$, $\quad$ Iwao SATO$^{5}$, $\quad$ Yuma TAMURA$^{6}$, 
\\ 
\\ 
\\
Department of Mathematical Sciences \\
College of Science and Engineering \\
Ritsumeikan University \\
1-1-1 Noji-higashi, Kusatsu, 525-8577, JAPAN \\
e-mail: akahori@se.ritsumei.ac.jp$^{1}$, \ horitakzk@gmail.com$^{2,\ast}$, \\
n-konno@fc.ritsumei.ac.jp$^{3}$, ra0099vs@ed.ritsumei.ac.jp$^{4}$, \\ ytamura11029@gmail.com$^{6}$ 
\\
\\
Oyama National College of Technology \\
771 Nakakuki, Oyama 323-0806, JAPAN \\
e-mail: isato@oyama-ct.ac.jp$^{5}$
}

\date{\empty }

\maketitle

\vspace{50mm}


\vspace{20mm}


\begin{small}
\par\noindent
{\bf Corresponding author}: Kazuki Horita, Department of Mathematical Sciences, College of Science and Engineering, Ritsumeikan University, 1-1-1 Noji-higashi, Kusatsu, 525-8577, JAPAN \\
e-mail: horitakzk@gmail.com
\par\noindent
\\
\par\noindent
{\bf Abbr. title: Grover algorithm and absolute zeta functions} 
\end{small}









\clearpage

\begin{abstract}
The Grover algorithm is one of the most famous quantum algorithms. On the other hand, the absolute zeta function can be regarded as a zeta function over $\mathbb{F}_{1}$ defined by a function satisfying the absolute automorphy. In this study, we show the property of the Grover algorithm and present a relation between the Grover algorithm and the absolute zeta function. We focus on the period of the Grover algorithm, because if the period is finite, then we are able to get an absolute zeta function explicitly by Kurokawa's theorem. In addition, whenever the period is finite or not, an expansion of the absolute zeta function can be obtained by a direct computation.
\end{abstract}

\vspace{10mm}

\begin{small}
\end{small}

\vspace{10mm}

\section{Introduction \label{sec01}}

This work is a continuation of our papers \cite{daiyon, AkahoriEtAL2023, dairoku, daiiti}. The origin of quantum computers was first proposed in 1980 by Benioff \cite{benioff} and the theory of quantum computer was established by Deutsch in 1985 \cite{Deutsch}. Quantum algorithms have also been studied as one of the main applications of quantum computers. Some theories of quantum algorithms were shown to perform as well as or better than classical algorithm on some problems. Shor's algorithm and Grover's algorithm are especially well-known. Shor's algorithm is for the problem of prime factorization, for which no efficient algorithm existed in classical computers \cite{Shor}. For an integer $N$, the algorithm efficiently finds its factors in a time which is polynomial in log $N$. Also, the Grover algorithm was proposed by Grover in 1996 \cite{Grover} and has been 
extensively studied by many researchers (see \cite{Portugal}, for example). The purpose of the algorithm is to find a certain item in unsorted $N$ times data. In classical algorithm, the most efficient algorithm for this problem is to examine the items one by one. Then, we need $O(N)$ steps in the classical algorithm. Moreover, the Grover algorithm is mathematically composed of the two unitary operators ${\cal R}_f$ and ${\cal R}_D$. Here, ${\cal R}_f$ plays the roles of inverting the sign and ${\cal R}_D$ diffuses uniformly. Now, the evolution operator $U$ performing one step of the Grover algorithm is given by product the two operators, i.e., $U = {\cal R}_f {\cal R}_D$. Then, we can obtain an interesting result that the algorithm can find in only $O(\sqrt{N})$ steps. This is because quantum computers can examine multiple items simultaneously due to the superposition of states. Furthermore, it is proved that the Grover algorithm is the fastest algorithm in quantum algorithm for the search problem.

On the other hand, the absolute zeta function is a zeta function on $\mathbb{F}_{1}$ defined by a function satisfying the absolute automorphy. Here, $\mathbb{F}_{1}$ is a limit of $\mathbb{F}_{p}$ as $p \rightarrow 1$, where $\mathbb{F}_{p} = \mathbb{Z}/p\mathbb{Z}$ for a prime number $p$. If we can find a time evolution matrix of a mathematical model that is an orthogonal matrix, then the matrix zeta function by the orthogonal matrix is absolute automorphy, thus the absolute zeta function is obtained. In our previous works, we obtained the absolute zeta functions of Grover walks, Hadamard walks, and more, see \cite{daiyon, AkahoriEtAL2023, dairoku, daiiti}.

In this study, we deal with the Grover algorithm. The $N \times N$ time-evolution matrix of the Grover algorithm, denoted by $U$, is an orthogonal matrix for $N=2,3,4, \dots$. Then we determine the period $T(U)$ of $U$. After that, we introduce the matrix zeta function $\zeta_{U}$ for $U$, which is an absolute automorphic form. Moreover, if $T(U) < \infty$ (i.e., $N=2,4$), then we got the absolute zeta function of $\zeta_{U}$, denoted by $\zeta_{ \zeta_{U}}$, and its functional equation by Kurokawa's theorem \cite{Kurokawa}. Additionally, for any $N=2,3,4, \dots$, an expansion of $\zeta_{ \zeta_{U}}$ can be derived. One of the motivations of our study is to connect the Grover algorithm with various branches of mathematics through absolute zeta function. That is, we will be able to apply some results in mathematics to the Grover algorithm. It would be helpful for quantum information processing to know a relation between Grover algorithm and mathematics, especially the absolute zeta function, by providing the first step for analyzing quantum algorithm.

The rest of this paper is organized as follows. Section 2 briefly describes the definition and property of the time-evolution operator $U$. In Section 3, we give the period of $U$ by two different methods. Section 4 explains the absolute zeta function and its related topics. In Section 5, we obtain absolute zeta functions and their functional equations. Finally, Section 6 is devoted to conclusion.



\section{Grover algorithm}

The Grover algorithm was proposed by Grover in 1996 \cite{Grover}. The algorithm is one of the most famous quantum algorithms, along with Shor's algorithm. The aim of the algorithm is to search an item in unsorted quantum database. It is known that the algorithm is optimal algorithm in quantum algorithm. 

In this section, we explain the Grover algorithm.
The dynamics of the Grover algorithm is described as a unitary matrix $U$ defined by product of two unitary matrices ${\cal R}_D$ and ${\cal R}_f $. Here, ${\cal R}_D$ plays the role of diffusing uniformly and ${\cal R}_f$ inverts the sign. By applying these two actions many times, you can find the desired item with a high probability. Then, it is well known that the algorithm requires a number of times on the order of $\sqrt{N}$ to the number $N$ to be searched. We assume that $N \in \mathbb{Z}_{>}$ with $N \geq 2 $, where $\mathbb{Z}_{>} = \{ 1, 2, 3, \cdots \}$.

First, let $I_N$ be the $N \times N$ identity matrix, and
\begin{align*}
\ket 0 = \  {}^T [1, 0, \ldots, 0],
\end{align*}
where $T$ means the transposed operator. For $N$ cards $\{0,1, \ldots, N-1\}$, ``$0$" is a marked card. In general, $\ket j$ is a vector whose $j+1$ component is 1 and the others are 0. Moreover, $| D \rangle $ is sometimes called the diagonal state, defined by
\begin{align}
| D \rangle = \frac{1}{\sqrt{N}} \  \sum_{j=0}^{N-1} | j \rangle = \frac{1}{\sqrt{N}} \  {}^T [1,1,\ldots,1]. \nonumber
\end{align}
It is easy to check that the norm $|| | D \rangle ||$ of the diagonal state $| D \rangle $ is equal to $1$.
The equation $\langle D | D \rangle  = 1$ is also used.
Furthermore, $| D \rangle \langle D |$ is a square matrix of order $N$ in which all the components are $1/N$. 

Now, an $N \times N$ matrix $U$ corresponding to the Grover algorithm is defined by
\begin{align*}
U={\cal R}_D {\cal R}_f, 
\end{align*}
where ${\cal R}_D$ and ${\cal R}_f$ are $N \times N$ matrices satisfying
\begin{align*}
{\cal R}_D &= 2 \ket D \bra D  - I_N, \\
{\cal R}_f &= I_N - 2 \ket 0 \bra 0.
\end{align*}

From now on, we call a matrix $U$ the Grover algorithm matrix. Then we have the following basic properties of ${\cal R}_D, {\cal R}_f$, and $U$.

\begin{pro}
${\cal R}_D, \  {\cal R}_f,$ and $U$ are unitary matrices.
\label{kinkyu01ch01}
\end{pro}
\par\noindent
{\bf Proof}. First, we prove that ${\cal R}_D$ is unitary. Because ${\cal R}_D$ is a real symmetric matrix, we see ${\cal R}_D^{\ast} = {\cal R}_D$. Noting $\bra D D \rangle = 1$, we have
\begin{align*}
{\cal R}_D {\cal R}_D^{\ast} 
&= {\cal R}_D^{\ast} {\cal R}_D = {\cal R}_D ^2 
\\
&= 
\left( 2 \ket D \bra D  - I_N \right) \left( 2 \ket D \bra D  - I_N \right) 
\\
&= 4 \ket D \bra D D \rangle \bra D - 4 \ket D \bra D + I_N = I_N.
\end{align*}
Similarly, we prove that ${\cal R}_f$ is a unitary matrix. Since ${\cal R}_f$ is also a real symmetric matrix, we get ${\cal R}_f^{\ast} = {\cal R}_f$. Thus, it follows from $\bra 0 0 \rangle = 1$ that
\begin{align*}
{\cal R}_f {\cal R}_f^{\ast} 
&= {\cal R}_f^{\ast} {\cal R}_f = {\cal R}_f ^2 
\\
&= 
\left( I_N - 2 \ket 0 \bra 0 \right) \left( I_N - 2 \ket 0 \bra 0 \right)  
\\
&= I_N  - 4 \ket 0 \bra 0 + 4 \ket 0 \bra 0 0 \rangle \bra 0 = I_N.
\end{align*}
Finally, we show the unitarity of $U$ by the unitarity of ${\cal R}_D$ and ${\cal R}_f$ in the following way.
\begin{align*}
U U^{\ast} 
&= ({\cal R}_D {\cal R}_f)({\cal R}_D {\cal R}_f)^{\ast} = {\cal R}_D {\cal R}_f{\cal R}_f ^{\ast} {\cal R}_D ^{\ast}
\\
&= {\cal R}_D I_N {\cal R}_D ^{\ast} = {\cal R}_D {\cal R}_D ^{\ast} = I_N.
\end{align*}

\par
\
\par
\section{Property of Grover algorithm \label{section012}}
In this section, we investigate properties of the unitary matrix $U$, i.e., the Grover algorithm matrix defined in the previous section. In order to do this, we also study ${\cal R}_D$ and ${\cal R}_f$.
First, we obtain the characteristic polynomials for  ${\cal R}_D, \  {\cal R}_f,$ and $ \  U$ as follows to consider the eigenvalues of ${\cal R}_D, \  {\cal R}_f,$ and $U$.
\begin{pro}
For any $N=2, \ 3, \ldots$, we have\
\begin{align}
\det \left( \lambda I_N -{\cal R}_D \right) = \left( \lambda - 1 \right) \left( \lambda + 1 \right)^{N-1},
\nonumber \\
\det \left( \lambda I_N -{\cal R}_f \right) = \left( \lambda - 1 \right)^{N-1} \left( \lambda + 1 \right), \nonumber \\
\det \left( \lambda I_N -U \right) = \left( \lambda + 1 \right)^{N-2} \left\{ \lambda^2 - 2 \left( 1 - \frac{2}{N} \right) \lambda + 1 \right\}. 
\label{shimuraken03}
\end{align}
\label{shimuraken}
\end{pro}
We should remark that the absolute value of the eigenvalues of ${\cal R}_D, \  {\cal R}_f,$ and $U$ are equal to $1$, because of the unitarity of  ${\cal R}_D, \  {\cal R}_f,$ and $U$.
Now, let ${\rm Spec} (A)$ be the set of the eigenvalues of a square matrix $A$. More precisely, we also use the notation:
\begin{align*}
    {\rm Spec} (A) = \{ [\lambda_1]^{m_1}, \ [\lambda_2]^{m_2}, \ \ldots, \ [\lambda_{\ell}]^{m_{\ell}}\},
\end{align*}
where $\lambda_{j}$ is the eigenvalue of $A$ and $m_{j} \in \mathbb{Z}_{>}$ is the multiplicity of $\lambda_{j}$ for $j = 1,2, \dots ,l$.
Then Proposition \ref{shimuraken} gives the following result.
\begin{cor}
For any $N=2,\ 3, \ldots$, we have
\begin{align*}
{\rm Spec} ({\cal R}_D) &= \{ [1]^1, \ [-1]^{N-1} \}, \\ 
{\rm Spec} ({\cal R}_f) &= \{ [1]^{N-1}, \ [-1]^{1} \}, \\
{\rm Spec} (U) &= \{ [-1]^{N-2}, \ [e^{i \xi}]^{1}, \  [e^{-i \xi}]^{1} \},
\end{align*}
where $\cos \xi= 1 - (2/N)$.
\label{stera}
\end{cor}

From now on, we consider $N=2,\ 3,$ and $4$ cases. For $N=2$ case, ${\cal R}_D, \  {\cal R}_f,$ and $U$ are given by
\begin{align*}
{\cal R}_D
=
\begin{bmatrix}
0&1\\
1&0
\end{bmatrix}
, \quad 
{\cal R}_f
=
\begin{bmatrix}
-1&0\\
0&1
\end{bmatrix}
,
\quad 
U={\cal R}_D {\cal R}_f
=
\begin{bmatrix}
0&1\\
-1&0
\end{bmatrix}
.
\end{align*}
Then, we can easily get the characteristic polynomials as follows.
\begin{align*}
\det \left( \lambda I_2 -{\cal R}_D \right) 
&= \det \left( \lambda I_2 -{\cal R}_f \right) = \left( \lambda - 1 \right) \left( \lambda + 1 \right),
\\
\det \left( \lambda I_2 -U \right) 
&= \lambda^2 + 1.
\end{align*}
Therefore, we have
\begin{align*}
{\rm Spec} ({\cal R}_D) 
&= {\rm Spec} ({\cal R}_f) = \{ [1]^1, [-1]^1 \}, \\
{\rm Spec} (U) 
&= \{ [i]^{1},  [-i]^{1} \}.
\end{align*}
Next, we consider the period of a matrix. The definition of the period of a matrix $A$ is given by
\begin{align*}
T= T(A)= \inf \left\{ t \ge 1 \  : \  A^t = I \right\}.
\end{align*}
If $\left\{ t \ge 1 \  : \  A^t = I \right\}$ is the empty set, then we define $T=\infty$.
By direct computation, we obtain 
\begin{align*}
{\cal R}_D ^2 = {\cal R}_f ^2 = I_2. 
\end{align*}
Thus, the period of ${\cal R}_D$ and ${\cal R}_f$ are $2$. On the other hand, $U^2=-I_2$ implies
\begin{align*}
U^4 = I_2 .
\end{align*}
Therefore, the period $T(U)$ is $4$. In summary, we see
\begin{align*}
T({\cal R}_D) = T({\cal R}_f) = 2, \quad T(U) = 4.
\end{align*}
In fact, the following result is useful to obtain the period.
\begin{pro}
Suppose that the eigenvalue of a unitary matrix $A$ is given as
\begin{align*}
{\rm Spec} (A) = \{ [\lambda_1]^{m_1}, \ [\lambda_2]^{m_2}, \ \ldots, \ [\lambda_{\ell}]^{m_{\ell}}\} 
\end{align*}
and $T < \infty$. Then,``a matrix $A$ has its period $T$" if and only if
\begin{align*}
T= \inf \left\{ t \ge 1 \  : \  \lambda_k ^t = 1 \quad (k=1,2, \ldots, \ell) \right\}.
\end{align*}
\label{coronashimura}
\end{pro}

It follows from Proposition \ref{shimuraken} that $T({\cal R}_D) = T({\cal R}_f) = 2$, since the eigenvalues of ${\cal R}_D$ and ${\cal R}_f$ are $1$ and $-1$. Similarly, we have $T(U) = 4$, because the eigenvalues of $U$ are $i$ and $-i$.

Next, we consider ${\cal R}_D, \  {\cal R}_f,$ and $U$ in $N=3$ case. From the definitions, we get
\begin{align*}
{\cal R}_D
= \frac{1}{3}
\begin{bmatrix}
-1&2&2\\
2&-1&2\\
2&2&-1
\end{bmatrix}
, \quad 
{\cal R}_f
=
\begin{bmatrix}
-1&0&0\\
0&1&0\\
0&0&1
\end{bmatrix}
,
\end{align*}
\begin{align*}
U={\cal R}_D {\cal R}_f
= \frac{1}{3}
\begin{bmatrix}
1&2&2 \\
-2&-1&2 \\
-2&2&-1 
\end{bmatrix}
.
\end{align*}
Then, we obtain
\begin{align}
\det \left( \lambda I_3 -{\cal R}_D \right) 
&= \left( \lambda - 1 \right) \left( \lambda + 1 \right)^2, \nonumber
\\
\det \left( \lambda I_3 -{\cal R}_f \right) 
&= \left( \lambda - 1 \right)^2 \left( \lambda + 1 \right), \nonumber
\\
\det \left( \lambda I_3 -U \right) 
&= \left( \lambda + 1 \right) \left( \lambda^2 - \frac{2}{3} \lambda + 1 \right). \nonumber
\end{align}
Therefore, we see
\begin{align*}
{\rm Spec} ({\cal R}_D) 
&= \{ [1]^1, \ [-1]^2 \}, 
\\
{\rm Spec} ({\cal R}_f) 
&= \{ [1]^2, \ [-1]^1 \}, 
\\
{\rm Spec} (U) 
&= \{[-1]^1, \  [e^{i \xi}]^{1}, \  [e^{-i \xi}]^{1} \} \quad \left( \cos \xi = 1/3 \right) .
\end{align*}
For ${\cal R}_D$ and ${\cal R}_f$, we obtain that the periods are 2 from the above calculation as in the case of $N=2$.
\begin{align*}
{\cal R}_D ^2 = {\cal R}_f ^2 = I_3 .
\end{align*}
On the other hand, it is difficult to check the period of $U$, because the identity matrix does not appear no matter how many times it is multiplied. In fact, a later discussion via cyclotomic polynomials will show that the period of $U$ is $\infty$. 
In conclusion, we obtain
\begin{align*}
T({\cal R}_D) = T({\cal R}_f) = 2, \quad T(U) = \infty.
\end{align*}
As in the case of $N=2$, Proposition \ref{coronashimura} confirms that the eigenvalues of ${\cal R}_D$ and ${\cal R}_f$ are $1$ and $-1$, so we obtain $T({\cal R}_D) = T({\cal R}_f) = 2$. However, $T(U) = \infty$ does not follow immediately from the eigenvalues of $U$.

We will consider ${\cal R}_D, \  {\cal R}_f$ and $ \  U$ in $N=4$ case. Then, ${\cal R}_D, {\cal R}_f$, and $U$ is given by
\begin{align*}
{\cal R}_D
= \frac{1}{2}
\begin{bmatrix}
-1&1&1&1\\
1&-1&1&1\\
1&1&-1&1\\
1&1&1&-1
\end{bmatrix}
, \quad 
{\cal R}_f
=
\begin{bmatrix}
-1&0&0&0\\
0&1&0&0\\
0&0&1&0\\
0&0&0&1
\end{bmatrix}
,
\end{align*}
\begin{align*}
U={\cal R}_D {\cal R}_f
= \frac{1}{2}
\begin{bmatrix}
1&1&1&1\\
-1&-1&1&1\\
-1&1&-1&1\\
-1&1&1&-1
\end{bmatrix}
.
\end{align*}
Thus, we see
\begin{align}
\det \left( \lambda I_4 -{\cal R}_D \right) 
&= \left( \lambda - 1 \right) \left( \lambda + 1 \right)^3, \nonumber
\\
\det \left( \lambda I_4 -{\cal R}_f \right) 
&= \left( \lambda - 1 \right)^3 \left( \lambda + 1 \right), \nonumber
\\
\det \left( \lambda I_4 -U \right) 
&= \left( \lambda + 1 \right)^2 \left( \lambda^2 - \lambda + 1 \right). \nonumber
\end{align}
Therefore, the eigenvalues are given by
\begin{align}
{\rm Spec} ({\cal R}_D) 
&= \{ [1]^1, \ [-1]^3 \},  \nonumber
\\
{\rm Spec} ({\cal R}_f) 
&= \{ [1]^3, \ [-1]^1 \}, \nonumber
\\
{\rm Spec} (U) 
&= \left\{ [-1]^2, \ \left[\frac{1+\sqrt{3}i}{2} \right]^{1}, \ \left[\frac{1-\sqrt{3}i}{2} \right]^{1} \right\}. \nonumber
\end{align}
A direct calculation implies
\begin{align*}
{\cal R}_D ^2 = {\cal R}_f ^2 = I_4.
\end{align*}
Thus, the periods of ${\cal R}_D$ and ${\cal R}_f$ are 2. On the other hand, we also get
\begin{align*}
U^6 = I_4.
\end{align*}
In conclusion, we see
\begin{align*}
T({\cal R}_D) = T({\cal R}_f) = 2, \quad T(U) = 6 .
\end{align*}

We considered $N=2, \ 3$, and $4$ cases. Based on these argument, we can derive the characteristic polynomials of Proposition \ref{shimuraken} in general $N$ case. Moreover, $U$ is a matrix reversed the sign of the first column of ${\cal R}_D$, where ${\cal R}_D$ is called the Grover matrix and plays an important role at various points in this paper. In general, the $(i,j)$ element $U_G (i,j)$ of the $N \times N$ Grover matrix $U_G$ is given as
\begin{align*} 
U_G (i,j) =
\begin{cases} 
\frac{2}{N} -1 & \text{$(i =j)$, } 
\\
\frac{2}{N} & \text{$(i \not =j)$}
\end{cases} 
\end{align*}
for $i,\ j =1, \ 2, \ldots , N$.

We investigate the periods of ${\cal R}_D, \  {\cal R}_f,$ and $  \  U$. First, we look at the period of ${\cal R}_D$ and ${\cal R}_f$. By Corollary \ref{stera}, we obtain that the eigenvalues of ${\cal R}_D$ and ${\cal R}_f$ are only $1$ and $-1$. In other words, $T({\cal R}_D)=T({\cal R}_f)=2$.

\par
In order to clarify the dependence on $N$, we write $U$ as $U_{N}$ and $T(U)$ as $T(U_{N})$. From now on, we find the period $T(U)$ for general $N \geq 2$. 

First, we recall that it follows from Eq. \eqref{shimuraken03} that the characteristic polynomials $\det (x I_N -U_{N})$ is given by
\begin{align}
\det \left( x I_N -U_{N} \right) = \left( x + 1 \right)^{N-2} \left\{ x^2 - 2 \left( 1 - \frac{2}{N} \right) x + 1 \right\} \quad (N=2,3, \ldots). 
\label{gralkihon}
\end{align}

The following result in Higuchi et al. \cite{HiguchiEtAl2017} is introduced.
\begin{pro}
    Let $f(x)$ be a monic polynomial with $\mathbb{Q}$-coefficients. If all solutions of $f(x)$ are roots of unity, then $f(x)$ can be uniquely factorized into cyclotomic polynomials. Hence $f(x)$ must be a monic polynomial with $\mathbb{Z}$-coefficients.
    \label{higuchi}
\end{pro}

Now, we use Proposition \ref{higuchi} to determine the period $T(U_{N})$ of $U_{N}$. Note that $T(U_{N}) < \infty$ if and only if all eigenvalues of $U_{N}$ are roots of unity. Obviously, Eq. \eqref{gralkihon} is a monic polynomial with $\mathbb{Q}$-coefficients. Therefore, if Eq. \eqref{gralkihon} is not a monic polynomials with $\mathbb{Z}$-coefficients, then $T(U_{N}) = \infty$. We easily see that Eq. \eqref{gralkihon} is a monic polynomial with $\mathbb{Z}$-coefficients, i.e., $2 \left( 1 - \frac{2}{N} \right) x \in \mathbb{Z}$ if and only if $N = 2$ and $4$.

If $N = 2$, then the solutions of $x^{2} + 1 = 0 $ are $i$ and $-i$. Thus, we get the conclusion $T(U_{2}) = 4$.

When $N = 4$, then the solutions of $x^{2} -x +1 = 0$ are $(1 + \sqrt{3}i) / 2$ and $(1 - \sqrt{3} i) / 2$. So, we obtain the period $T(U_{4}) = 6$.

Therefore, we obtain the following result on the period of $U_{N}$.
\begin{thm}
Let $U_{N}$ be the Grover algorithm matrix with $N=2,\ 3, \ldots$. Then the period $T(U_{N})$ is given by
\begin{align*} 
T(U_{N})
=
\begin{cases} 
4 & \text{$(N=2)$, } 
\\
6 & \text{$(N=4)$,} 
\\ 
\infty & \text{$(N \not=2, \ 4)$.} 
\end{cases} 
\end{align*}
\label{gralperi}
\end{thm}
We should remark that $T(U_{N})$ depends on $N$ unlike the case of $R_{D}$ and $R_{f}$. In fact, $T(R_{D}) = T(R_{f}) = 2$ for any $N \geq 2$.
This result is also proved using the cyclotomic polynomial.

From now on, we obtain Theorem \ref{gralperi} by using the cyclotomic polynomial. To do so, we introduce cyclotomic polynomials.
\begin{deff}
    For $n \in \mathbb{Z}_{>}$, the cyclotomic polynomial $\Phi_{n} (x)$ is defined by
    \begin{align*}
        \Phi_{n} (x) = \prod_{\substack{1 \leq k \leq n-1 \\ \mathrm{gcd}(k,n) = 1}} \left( x - \mathrm{exp} \left(\frac{2 \pi i k}{n} \right) \right).
    \end{align*}
\end{deff}
It is well known that there are only 3 types of cyclotomic polynomials whose degree is $2$ as follows.
\begin{align*}
\Phi_3 (x) = x^2+x+1, \quad \Phi_4 (x) = x^2+1, \quad \Phi_6 (x) = x^2-x+1. 
\end{align*}
Thus, if the period $T(U_{N})$ is finite, then the corresponding quadratic polynomial part of the characteristic polynomial $\det (x I_N -U_{N})$, that is,
\begin{align*}
x^2 - 2 \left( 1 - \frac{2}{N} \right) x + 1
\end{align*}
is equal to  $\Phi_3 (x), \ \Phi_4 (x), $ or $ \Phi_6 (x)$. Therefore we see
\begin{align*}
- 2 \left( 1 - \frac{2}{N} \right) \in \{ 0, \pm 1 \}.
\end{align*}
From now on, we consider the each case.

\par
First, if $- 2 \left( 1 - \frac{2}{N} \right) = 1$, then $N=4/3$, so there does not exist such an integer $N$.

\par
Next, when $- 2 \left( 1 - \frac{2}{N} \right) = 0$, we have $N=2$. Then the quadratic polynomial part equals to $\Phi_4 (x) = x^2+1$, i.e., $T(U_{2})=4$. In fact,
\begin{align*}
{\rm Spec} (U_{2}) = \{ [i]^{1}, \  [-i]^{1} \} ,
\end{align*}
thus we get the same conclusion, i.e., $T(U_{2})=4$.

\par
Finally, if $- 2 \left( 1 - \frac{2}{N} \right) = -1$, we take $N=4$. Then the quadratic polynomial part is equal to $\Phi_6 (x) = x^2-x+1$. Therefore, we get $T(U_{4}) = 6$. Indeed, we have
\begin{align*}
{\rm Spec} (U_{4}) = \left\{ [-1]^2, \ \left[\frac{1+\sqrt{3}i}{2} \right]^{1}, \ \left[\frac{1-\sqrt{3}i}{2} \right]^{1} \right\} .
\end{align*}
Therefore, we obtained the same conclusion in Theorem \ref{gralperi} by the cyclotomic polynomial.

\section{Absolute zeta function} 
In this section, we briefly review the framework on the absolute zeta functions, which can be considered as a zeta function over $\mathbb{F}_1$, and absolute automorphic forms (see \cite{Kurokawa3, Kurokawa, KO, KT3, KT4} and references therein, for example). 

Let $f(x)$ be a function $f : \mathbb{R} \to \mathbb{C} \cup \{ \infty \}$. We say that $f$ is an {\em absolute automorphic form} of weight $D$ if $f$ satisfies
\begin{align*}
f \left( \frac{1}{x} \right) = C x^{-D} f(x)
\end{align*}
with $C \in \{ -1, 1 \}$ and $D \in \mathbb{Z}$. The {\em absolute Hurwitz zeta function} $Z_f (w,s)$ is defined by
\begin{align*}
Z_f (w,s) = \frac{1}{\Gamma (w)} \int_{1}^{\infty} f(x) \ x^{-s-1} \left( \log x \right)^{w-1} dx,
\end{align*}
where $\Gamma (x)$ is the gamma function (see \cite{Andrews1999}, for instance). Then, taking $x=e^t$, we see that $Z_f (w,s)$ can be rewritten as the Mellin transform: 
\begin{align}
Z_f (w,s) = \frac{1}{\Gamma (w)} \int_{0}^{\infty} f(e^t) \ e^{-st} \ t^{w-1} dt. \nonumber
\end{align}
Moreover, the {\em absolute zeta function} $\zeta_f (s)$ is defined by 
\begin{align*}
\zeta_f (s) = \exp \left( \frac{\partial}{\partial w} Z_f (w,s) \Big|_{w=0} \right).
\end{align*}
Here, we introduce the {\em multiple Hurwitz zeta function of order $r$}, $\zeta_r (s, x, (\omega_1, \ldots, \omega_r))$, the {\em multiple gamma function of order $r$}, $\Gamma_r (x, (\omega_1, \ldots, \omega_r))$, and the {\em multiple sine function of order $r$}, $S_r (x, (\omega_1, \ldots, \omega_r))$, respectively (see \cite{Kurokawa3, Kurokawa, KT3}, for example): 
\begin{align}
\zeta_r (s, x, (\omega_1, \ldots, \omega_r))
&= \sum_{n_1=0}^{\infty} \cdots \sum_{n_r=0}^{\infty} \left( n_1 \omega_1 + \cdots + n_r \omega_r + x \right)^{-s}, \nonumber
\\
\Gamma_r (x, (\omega_1, \ldots, \omega_r)) 
&= \exp \left( \frac{\partial}{\partial s} \zeta_r (s, x, (\omega_1, \ldots, \omega_r)) \Big|_{s=0} \right), \nonumber
\\
S_r (x, (\omega_1, \ldots, \omega_r))
&= \Gamma_r (x, (\omega_1, \ldots, \omega_r))^{-1} \ \Gamma_r (\omega_1+ \cdots + \omega_r - x, (\omega_1, \ldots, \omega_r))^{(-1)^r}. \nonumber
\end{align}
\par
Now we present the following key result derived from Theorem 4.2 and its proof in Korokawa \cite{Kurokawa} (see also Theorem 1 in Kurokawa and Tanaka \cite{KT3}):

\vspace*{12pt}
\noindent
\begin{thm}
For $\ell \in \mathbb{Z}, \ m(i) \in \mathbb{Z}_{>} \ (i=1, \ldots, a), \ n(j) \in \mathbb{Z}_{>} \ (j=1, \ldots, b)$, put 
\begin{align*}
f(x) = x^{\ell/2} \ \frac{\left( x^{m(1)} - 1 \right) \cdots \left( x^{m(a)} - 1 \right)}{\left( x^{n(1)} - 1 \right) \cdots \left( x^{n(b)} - 1 \right)}.
\end{align*}
Then we have 
\begin{align}
Z_f (w, s) 
&= \sum_{I \subset \{1, \ldots, a \}} (-1)^{|I|} \ \zeta_b \left( w, s - {\rm deg} (f) + m \left( I \right), \bec{n} \right), \nonumber
\\
\zeta_f (s) 
&= \prod_{I \subset \{1, \ldots, a \}} \Gamma_b \left( s - {\rm deg} (f) + m \left( I \right), \bec{n} \right)^{ (-1)^{|I|}}, \nonumber
\\
\zeta_f \left( D-s \right)^C 
&= \varepsilon_f (s) \ \zeta_f (s), 
\label{mkusatsufe}
\end{align}
where
\begin{align*}
|I| 
&= \sum_{i \in I} 1, \quad {\rm deg} (f) = \frac{\ell}{2} + \sum_{i=1}^a m(i)- \sum_{j=1}^b n(j), \quad m \left( I \right) = \sum_{i \in I} m(i),
\\
\bec{n} 
&= \left( n(1), \ldots, n(b) \right), \quad D = \ell + \sum_{i=1}^a m(i)- \sum_{j=1}^b n(j), \quad C=(-1)^{a-b}, 
\\
\varepsilon_f (s) 
&= \prod_{I \subset \{1, \ldots, a \}} S_b \left( s - {\rm deg} (f) + m \left( I \right), \bec{n} \right)^{ (-1)^{|I|}}.
\end{align*}
\label{zettaisugaku01} 
\end{thm}
\vspace*{12pt}
\noindent
We should note that Eq. \eqref{mkusatsufe} is called the {\em functional equation}.
\section{Grover algorithm and abusolute zeta function \label{groverabso}}

In our previous works, we considered the various absolute zeta functions of time-evolution matrices related to quantum models, for example, quantum walks \cite{daiyon, daiiti}, quantum cellular automata \cite{AkahoriEtAL2023}, and quantum version of Markov chains \cite{dairoku}. In each case, we considered the matrix zeta function and its absolute zeta function. In other words, we obtained the absolute zeta function of a zeta function.  

In this section, the size of the Grover algoritm matrix $U$ is important, so we also write the $N \times N$ matrix $U$ as $U_N$ as in the last part of Section 3. It follows from Theorem \ref{gralperi} that $T(U_{N})$ is finite for $N = 2$ and $4$. So we will calculate absolute zeta functions for both cases by Theorem \ref{zettaisugaku01}. We start with the basic equation, i.e., Eq. \eqref{gralkihon}:
\begin{align}
\det \left( x I_N - U_N \right) = \left( x + 1 \right)^{N-2} \left\{ x^2 - 2 \left( 1 - \frac{2}{N} \right) x + 1 \right\} .
\label{gralkihonb}
\end{align}
Then putting $x = 1/u$ implies 
\begin{align}
\det \left( I_N - u U_N \right) 
&= u^N \cdot \det \left( \frac{1}{u} I_N - U_N \right) \nonumber
\\
&= u^N \cdot \left( \frac{1}{u} + 1 \right)^{N-2} \left\{ \left( \frac{1}{u} \right)^2 - 2 \left( 1 - \frac{2}{N} \right) \left( \frac{1}{u} \right) + 1 \right\} \nonumber
\\
&= \left( 1 + u \right)^{N-2} \left\{ 1 - 2 \left( 1 - \frac{2}{N} \right) u  + u^2 \right\}.
\label{groverzetafunction}
\end{align}
Note that the second equality comes from Eq. \eqref{gralkihonb}.
\begin{deff}
The matrix zeta function $\zeta_{U_N} (u)$ of the Grover algorithm matrix $U_N$ is defined as
\begin{align*}
\zeta_{U_N} (u)^{-1} = \det \left( I_N - u U_N \right).
\end{align*}
\label{corgralkihonb}
\end{deff}
Then Eq. \eqref{groverzetafunction} gives
\begin{pro}
\begin{align*}
    \zeta_{U_N} (u)^{-1} = \left( u + 1 \right)^{N-2} \left\{ u^2 - 2 \left( 1 - \frac{2}{N} \right) u  + 1 \right\}.
\end{align*}
\label{groverzeta02}
\end{pro}
From the definition of $\zeta_{U_{N}}(u)$, we have the following key result.
\begin{pro}
\begin{align*}
\zeta_{U_N} \left( \frac{1}{u} \right) = (-1)^{N} \ \det (U_N) \ u^{N} \ \zeta_{U_N} \left( u \right).
\end{align*}
\label{kaginocorisec01ga}
\end{pro}
We should remark that $\zeta_{U_N} (u)$ is an absolute automorphic form with weight $-N$, since the unitarity of $U_{N}$ implies $(-1)^{N} \det (U_N)  \in  \{ -1, 1 \}$.

\subsection{$N=2$ case}
This section deals with $N=2$ case.
From Proposition \ref{groverzeta02}, we get
\begin{align*}
\zeta_{U_2} (u)^{-1} = \det \left( I_2 - u U_2 \right) = \left( u + 1 \right)^{2-2} \left\{ u^2 - 2 \left( 1 - \frac{2}{2} \right) u  + 1 \right\} = u^2 + 1 .
\end{align*}
Therefore,
\begin{align*}
\zeta_{U_2} (u)^{-1} = u^2 + 1 = \Phi_4 (u) = \frac{u^4-1}{u^2-1} .
\end{align*}
Thus,
\begin{align*}
\zeta_{U_2} (u) = \frac{u^2-1}{u^4-1} .
\end{align*}
Then the above equation implies
\begin{align*}
\zeta_{U_2} \left( \frac{1}{u} \right) = u^{2} \ \zeta_{U_2} (u),
\end{align*}
Therefore $\zeta_{U_2} (u)$ is an absolute automorphic form with weight $-2$. Moreover, we obtain the following results by Theorem \ref{zettaisugaku01}.
\begin{align}
Z_{\zeta_{U_2}} (w, s) 
&= \zeta_1 \left( w, s + 2, (4) \right) - \zeta_1 \left( w, s + 4, (4) \right), \nonumber
\\
\zeta_{\zeta_{U_2}} (s)
&= \frac{\Gamma_1 \left( s+2, (4) \right)}{\Gamma_1 \left( s+4, (4) \right)}, \nonumber
\\
\zeta_{\zeta_{U_2}} (-2-s)
&= \varepsilon_{\zeta_{U_{2}}} (s) \ \zeta_{\zeta_{U_2}} (s), 
\label{groverzetanga2}
\end{align}
where $\ell =0, \ a=1, \ m(1)=2, \ b=1, \ n(1)=4, \ {\rm deg} (f) = D = -2, \  C=1$, and 
\begin{align}
\varepsilon_{\zeta_{U_{2}}} (s) \nonumber &=  \prod_{ I \subset \{ 1 \}} S_{1} \left( s-(-2)+m(I), (4) \right) ^{(-1)^{|I|}} \\ \nonumber
&= S_{1} \left( s- (-2) +2, (4) \right) ^{(-1)^{1}} \times S_{1} \left( s-(-2) +0 , (4) \right) ^{(-1)^{0}}  \\ \nonumber
&=S_{1} \left( s+4,(4) \right) ^{-1} S_{1} \left( s+2,(4) \right) ^{1} \\ \nonumber
&=\frac{S_{1} \left( s+2,(4) \right)}{S_{1} \left( s+4,(4) \right)} 
= -\frac{S_{1} \left( s+2,(4) \right)}{S_{1} \left( s,(4) \right)} 
= - \cot \left( \frac{s \pi}{4} \right). \nonumber
\end{align}
Note that the fifth equality comes from $S_{1} (x+\omega , (\omega) ) = -S_{1}(x , (\omega) )$. In addition, the sixth equality is given by $S_{1} (x, (\omega) ) = 2 \sin \left( \frac{\pi x }{\omega } \right)$.
Then, from the above functional equation, i.e., Eq. \eqref{groverzetanga2}, we take the central point: $s = -1$, since $-2-s = s$. Indeed, we have the central critical value at the central point $s= -1$:
\begin{align*}
\zeta_{\zeta_{U_2}} (-1) = \frac{\Gamma_{1} \left( 1, (4) \right)}{\Gamma_{1} \left( 3, (4) \right)} = \frac{\Gamma \left( \frac{1}{4} \right)}{2 \Gamma \left( \frac{3}{4} \right)}.
\end{align*}
Remark that the second equality comes from
\begin{align*}
\Gamma_{1} \left( x, (\omega) \right) = \frac{1}{\sqrt{2 \pi}} \Gamma \left( \frac{x}{\omega} \right) \omega^{\frac{x}{\omega} - \frac{1}{2}}.
\end{align*}

\subsection{$N=4$ case}
In this section, we treat $N=4$ case.
By Eq. \eqref{gralkihonb}, we see
\begin{align*}
\zeta_{U_4} (u)^{-1} 
&= \det \left( I_4 - u U_4 \right) = \left( u + 1 \right)^{4-2} \left\{ u^2 - 2 \left( 1 - \frac{2}{4} \right) u  + 1 \right\} 
\\
&= (u+1)^2 (u^2 - u + 1).
\end{align*}
Therefore, we have
\begin{align*}
\zeta_{U_4} (u)^{-1} 
&= (u+1)^2 (u^2 - u + 1) = \Phi_2 (u)^2 \Phi_6 (u) 
\\
&= \frac{(u^2-1)^2}{(u-1)^2} \cdot \frac{(u-1)(u^6-1)}{(u^2-1)(u^3-1)} = \frac{(u^2-1)(u^6-1)}{(u-1)(u^3-1)}.
\end{align*}
Thus,
\begin{align*}
\zeta_{U_4} (u) = \frac{(u-1)(u^3-1)}{(u^2-1)(u^6-1)}.
\end{align*}
From the above equation, $\zeta_{U_4} (u)$ satisfies
\begin{align*}
\zeta_{U_4} \left( \frac{1}{u} \right) = u^{4} \ \zeta_{U_4} (u).
\end{align*}
Therefore $\zeta_{U_4} (u)$ is an absolute automorphic form with weight $-4$. Furthermore, it follows from Theorem \ref{zettaisugaku01} that
\begin{align*}
Z_{\zeta_{U_4}} (w, s) 
&= \sum_{I \subset \{1,2 \}} (-1)^{|I|} \ \zeta_{2} \left(w, s + 4  + m(I), (2,6) \right),
\\
\zeta_{\zeta_{U_4}} (s)
&= \prod_{I \subset \{1,2 \}} \Gamma_{2} \left( s + 4  + m(I), (2,6) \right)^{ (-1)^{|I|}},
\\
\zeta_{\zeta_{U_4}} (-4-s) 
&= \varepsilon_{\zeta_{U_4}} (s) \ \zeta_{\zeta_{U_4}} (s),
\end{align*}
where $\ell =0, \ a=2, \ m(1)=1, \ m(2)=3, \ b=2, \ n(1)=2, \ n(2)=6, \ {\rm deg} (f) = D = -4, \  C=1$, and 
\begin{align*}
\varepsilon_{\zeta_{U_4}} (s) 
&= \prod_{I \subset \{1,2 \}} S_{2} \left( s + 4  + m(I), (2,6) \right)^{ (-1)^{|I|}}.
\end{align*}

\subsection{General $N$ case}
Finally, we consider the absolute zeta function of $U_{N}$ for any $N \in \mathbb{Z}_{>}$ with $N \geq 2$. Remark that for $N \neq 2,4$, we are not able to apply Theorem \ref{zettaisugaku01}.

We begin with
\begin{align*}
\zeta_{U_N} (u) 
&= \det \left( I_N - u U_N \right)^{-1} = \frac{1}{\left( 1 + u \right)^{N-2}} \ \frac{1}{1 - 2 \left( 1 - \frac{2}{N} \right) u  + u^2}
\\
&= \frac{\left( 1 - u \right)^{N-2}}{\left( 1 - u^2 \right)^{N-2}} \ \frac{1}{u^2 \left\{ 1 - 2 \left( 1 - \frac{2}{N} \right) \frac{1}{u}  + \frac{1}{u^2} \right\}}
\\
&= \left( 1 - u \right)^{N-2} \times (-1)^{N-2} \left\{ u^{-2} \sum_{r=0}^{\infty} \left( \frac{1}{u} \right)^{2r} \right\}^{N-2} \times \left( \sum_{\ell = 0}^{\infty} P_{\ell} \ u^{-\ell -2} \right)
\\
&= (-1)^{N-2} \sum_{m=0}^{N-2} {\binom{N-2}{m}} (-1)^m u^m \left( \sum_{r=0}^{\infty} u^{-2r-2} \right)^{N-2} \left(\sum_{\ell=0}^{\infty} P_{\ell} \ u^{-\ell-2} \right)
\\
&= (-1)^{N-2} \sum_{m=0}^{N-2} \sum_{r_1=0,\ldots,r_{N-2}=0}^{\infty} \sum_{\ell=0}^{\infty} {\binom{N-2}{m}} (-1)^m  P_{\ell} \ u^{m - 2 \sum_{q=1}^{N-2}{(r_{q}+1)}-(\ell+2)}.
\end{align*}
Let $\alpha$ and $\beta$ be solutions of a quadratic equation $u^2 -2(1-(2/N))u + 1=0$. Now, we use the notation $P_{\ell}=\sum_{k,r} \alpha^{k}\beta^{r}$. Moreover, for $\ell =0, 1, 2, \ldots$, we have
\begin{align*}
P_{\ell}=\sum_{k,r} \alpha^{k}\beta^{r}=\sum_{i=0}^{\lfloor \frac{\ell}{2} \rfloor}(-1)^i { \binom{\ell-i}{i} } (\alpha \beta)^i (\alpha + \beta)^{\ell-2i}=\sum_{i=0}^{\lfloor \frac{\ell}{2} \rfloor}(-1)^{i} { \binom{\ell - i}{i} } \left\{ 2 \left( 1 - \frac{2}{N} \right) \right\}^{\ell-2i},
\end{align*}
where $\mathbb{Z}_{\ge} = \{0,1,2, \ldots \}$ and $\sum_{k,r} $ means $\sum_{\{(k,r) \in (\mathbb{Z}_{\ge})^2 \ : \ k+r=\ell \}}$.

Then, we can cancel the gamma function $\Gamma (w)$ to transform the series expansion for $u$ as above, when we calculate the absolute Hurwitz zeta function. Thus, we obtain
\begin{align*}
Z_{\zeta_{U_{N}}}(w,s)
&= \frac{(-1)^{N-2}}{\Gamma(w)}\int_{1}^{\infty}\sum \Biggl( (-1)^m P_{\ell} \Biggr) x^{m - 2 \sum_{q=1}^{N-2} (r_{q}+1) - (\ell +2) -s-1} (\log x)^{w-1} dx 
\\
&=\frac{(-1)^{N-2}}{\Gamma(w)}\int_{0}^{\infty}\sum \Biggl( (-1)^m P_{\ell} \Biggr) e^{- \{ 2 \sum_{q=1}^{N-2} (r_{q}+1) + (\ell +2) - m + s \} t} \ t^{w-1}dt
\\
&= (-1)^{N-2} \sum \Biggl( (-1)^m P_{\ell} \Biggr) \frac{1}{\Gamma(w)}\int_{0}^{\infty} e^{ - \{ 2 \sum_{q=1}^{N-2} (r_{q}+1) + (\ell +2) - m + s \} t} \ t^{w-1} \ dt 
\\
&= (-1)^{N-2} \sum \Biggl( (-1)^m P_{\ell} \Biggr) \frac{1}{\{ 2 \sum_{q=1}^{N-2} (r_{q}+1) + (\ell +2) - m + s \}^w}.  
\end{align*}
Here, $\sum$ is short for $\sum_{m=0}^{\infty} \sum_{r_1=0,\ldots,r_{N-2}=0}^{\infty} \sum_{\ell=0}^{\infty}$. In addition, if $L$ is large enough, then
\begin{align*}
\sum \left\{ 2 \sum_{q=1}^{N-2} (r_{q}+1) + (\ell +2) - m + s \right\}^{-w}
\end{align*}
converges for $w$ in the range of $\Re(w) > L$, where $\Re(z)$ is the real part of $z \in \mathbb{C}$.
Here, $w$ is defined in the range of $\Re (w) > L$. By the analytic continuation from the area of $\Re (w) > L$, we can calculate the absolute zeta function $\zeta_{U_{N}}$ as follows.
\begin{align*}
\zeta_{\zeta_{U_{N}}}(s) 
&= \exp \biggl[ \frac{\partial}{\partial w} (-1)^{N-2} \sum \bigl((-1)^m P_{\ell} \bigr) \frac{1}{\left\{ 2 \sum_{q=1}^{N-2} (r_{q}+1) + (\ell +2) - m + s  \right\}^w} \Big|_{w=0} \biggr]
\\
&= \exp \biggl[ \sum (-1)^{N-2} \Bigl( (-1)^m P_{\ell} \Bigr) \log \biggl\{ \frac{1}{2 \sum_{q=1}^{N-2} (r_{q}+1) + (\ell +2) - m + s} \biggr\} \biggr]
\\
&= \exp \biggl[ \sum \log \biggl\{ 2 \sum_{q=1}^{N-2} (r_{q}+1) + (\ell +2) - m + s \biggr\}^{(-1)^{N+m} P_{\ell}} \biggr]
\\
&= \prod_{} \biggl\{ 2 \sum_{q=1}^{N-2} (r_{q}+1) + (\ell +2) - m + s \biggr\}^{(-1)^{N+m} P_{\ell}},
\end{align*}
where $\prod$ is an abbreviation of $\prod_{m=0}^{\infty} \prod_{r_1=0,\ldots,r_{N-2}=0}^{\infty} \prod_{\ell=0}^{\infty}$. 
\begin{thm}
    For $N \in \mathbb{Z}_{>}$ with $N \geq 2$, we have an expansion of the absolute zeta function:
    \begin{align*}
        \zeta_{\zeta_{U_{N}}}(s) = \prod_{} \biggl\{ 2 \sum_{q=1}^{N-2} (r_{q}+1) + (\ell +2) - m + s \biggr\}^{(-1)^{N+m} P_{\ell}}.
    \end{align*}
    \label{generalzettaizeta}
\end{thm}

Here we should remark a relation between our result and the following result by Kurokawa \cite{Kurokawa}.
\begin{pro}
We put $f_{\bec{\omega}}(x) = \left\{(1-x^{-\omega_1}) (1-x^{-\omega_2}) \cdots (1-x^{-\omega_r}) \right\}^{-1}$, where $\bec{\omega} = (\omega_1,\ldots, \omega_r),$  $\ \omega_1,\ldots,\omega_r>0$. Then, the absolute zeta function of $f_{\bec{\omega}}$ is given by
\begin{align}
\zeta_{f_{\bec{\omega}}} (s) = \Gamma_{r}(s,\bec{\omega})=\regprod_{\bec{n} \geq \bec{0}}(\bec{n} \cdot \bec{\omega} + s)^{-1} = \regprod_{n_1=0,\ldots,n_r=0}^{\infty} \left( n_1 \omega_1 + \cdots +  n_r \omega_r + s \right)^{-1}.
\label{arere}
\end{align}
\label{generalprop}
\end{pro}
That is, the absolute zeta function $\zeta_{\zeta_{U_{N}}}(s)$ for the Grover algorithm in Theorem \ref{generalzettaizeta} is an analogue of the above absolute zeta function $\zeta_{f_{\bec{\omega}}} (s)$ in Proposition \ref{generalprop}, where $\regprod$ in Eq. \eqref{arere} is the zeta regularized product.

\section{Conclusion}
In this paper, we considered the Grover algorithm $U_{N}$ for $N \in \mathbb{Z}_{>}$ with $N \geq 2$. First, we obtained the period $T(U_{N})$ in Theorem \ref{gralperi}: $T(U_{N}) = 4$ $(N=2)$, $=6$ $(N=4)$, and $= \infty$ $(N \neq 2,4)$. Then we gave two different proofs, one comes from Proposition \ref{higuchi}, and the other is based on cyclotomic polynomials. Afterwards, we introduced the matrix zeta function $\zeta_{U_{N}}$ for $U_{N}$, which is an absolute automorphic form with weight $-N$ (Proposition \ref{kaginocorisec01ga}). Furthermore, if $T(U_{N}) < \infty $ (i.e., $N=2,4$), then we got the absolute zeta function of $\zeta_{U_{N}}$, that is, $\zeta_{ \zeta_{U_{N}}}$ and its functional equation by Theorem \ref{zettaisugaku01} given in Kurokawa \cite{Kurokawa, KT3}. In addition, for any $N \in \mathbb{Z}_{>}$ with $N \geq 2$, an expansion of $\zeta_{ \zeta_{U_{N}}}$ was shown (Theorem \ref{generalzettaizeta}) which is an analogous result to Proposition \ref{generalprop} by Kurokawa \cite{Kurokawa}.

\par
\
\par
\noindent

\section*{Author Contributions}
The authors have no conflicts to disclose.

\section*{Data Availability}
Not applicable.

\section*{Conflict of Interest}
Not applicable.

\par
\
\par
\noindent


\end{document}